# Design and fabrication of low cost Disdrometer and turbulence sensor.

Amit Ram Morarka[1, a], Subhash Vasant Ghaisas[2,b]

[1]Department of Electronics Science, Savitribai Phule Pune University, Pune-411007, INDIA

[2]ME & MS department, Indian Institute of Technology, Mumbai, INDIA

[a]amitmorarka@gmail.com, [b]svgunipune@gmail.com

**Abstract.** Disdrometers detects size and speed of falling rain droplets in a rainfall. They detect droplets either by impact or optical based methods. Optical based ones, use light source and an optical detector. Droplet crossing the light creates shadow on the detector. This change in the intensity and duration of change provides the size and speed of the droplet. Whereas the impact based method utilizes peizo elements to sense the droplet by physical momentum transfer at the time of droplet's impact with the sensor. Apart from the technical differences in them, both of these disdrometers fall in the similar cost range. Wind turbulence sensors works on detection of gradients in wind speeds by employing various working principles like thermal anemometer, laser Doppler anemometer, etc. In thermo anemometers, a heating element is maintained at a constant temperature. As the wind flows, it carries some heat energy away from the element. This causes a change in the resistance in the heating element providing further change in voltages across it. Laser Doppler anemometers work on change in wavelength of light scattered by a moving particle which is caught in the wind flow. These techniques are highly accurate in measuring the turbulence in the air but at elevated costs. Aimed at cost efficacy and acceptable performance of the device, a novel concept was implemented in a single device to study its feasibility to work as a disdrometer and a turbulence sensor. It employs the principles of electromagnetic induction and response of a cantilever to impulses. Based on these principles, a device was designed and fabricated to detect rain droplet size and wind turbulence. It was fabricated from readily available materials. The testing of the device in laboratory conditions yielded satisfactory results as compare to the standard commercial instruments.

**Introduction**

Droplet size distribution is one of the most significant studies that are needed for the understanding of how rainfall affects the regional ecosystem. Droplet size estimation provides us the knowledge of how the rainfall intensity will cause the erosion in the soil. To determine the size of the rainfall droplet the instrument is called as Disdrometer. Many different disdrometers are available based on their working principles, cost and accuracy. One such optical based disdrometer is reported by [1]. They produce a laser light sheet from a 780nm laser diode. The sensor sees this light sheet and produces 5V signal. As a droplet passes through the sheet making a shadow over the sensor, the output voltage is also decreased in proportion with the size of the droplet. With this sensor not only the size but also velocity of the droplet can be determined. Their instrument can detect diameter of the droplets in the range of 0.3mm to 30mm. Another type of disdrometers is of impact based. Their sub classes are acoustic [2] and displacement based [3] disdrometers respectively. Electrical signals are generated by the piezoelectric sensor when a droplet makes an impact with it. Joss-Waldvogel Disdrometer the other sub class which works on displacement of a cone shaped diaphragm to produce magnetic induction in a coil attached to it. It is one of the most prominently used all over the world for the study of droplet size distribution. Both of them could measure reliably raindrop size having diameter in the range 1-6mm respectively. On the other hand wind turbulence sensors have a very diverse spread of techniques employed to detect it. Two dimensional atmospheric laser cantilever anemometer [4] works on the principle similar to that of an atomic force microscope (AFM). The drag force generated due to the flow of air on the cantilever makes it move causing a displacement in it. A mirror mounted on this cantilever reflects a light on to a Position sensitive detector which tracks these changes and outputs the data. It has a resolution in the range of millimeters. Hot wire anemometers works on the principle of sensing changes in current due to loss in temperature to the surrounding fluid flow as reported by [5-6]. Their major drawback is that their accuracy dependents greatly upon the fluid temperature. Looking at such diverse ways of sensing droplets and wind turbulence and their cost efficacy, a novel idea was conceptualized. When a raindrop impacts on to a plant leaf, it undergoes oscillation. The same leaf vibrates when wind flows by it. With this motivation it was thought that an electromagnetic induction and one end fixed cantilever based sensor can be designed and fabricated which could be useful to detect the size of the rain droplets and also will have capability to measure turbulence in the wind flows. To the best of our knowledge till date there are no reports of development of such hybrid and low cost sensor. In this report we describe feasibility of fabrication and usefulness of the hybrid sensor based on the principles of electromagnetic induction and response of a cantilever to impulses.

**Experimental**

Figure 1 shows schematics and the photo image of the developed hybrid sensor.

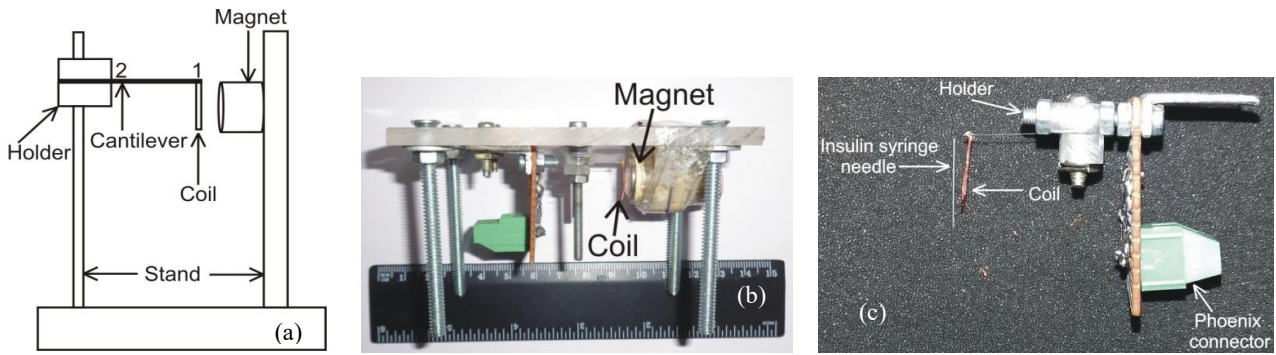

**Figure 1(a):** It shows Schematic of the sensor. (b) Side view image of the sensor alongwith its housing. (c) Side view image of the standalone cantilever-coil assembly of the sensor.

As shown in figure 1, the sensor works on principle of electromagnetic induction and cantilever. It has a similarity with Joss-Waldvogel disdrometer's sensing principle of electromagnetic induction. The novelty in our hybrid sensor system is that the same sensor without undergoing any geometrical or material changes can also detect turbulence present in the wind. While designing this sensing element one important aspect was taken into account was that the magnitude of the momentum transfer from rain droplet and the wind flows is of miniscule amount. It has to be transferred as efficiently as possible. Taking into account all the physical conditions, the cost efficacy and the technical aspects, a simple technique was employed. A coil was made from 17µm diameter enamel copper wire on an in-house developed wire winding machine. The specifications of the coil are given below in the table-1.

**Table-01:** Technical specifications of the coil

| Sr. no. | Description | Value | Unit |
|---|---|---|---|
| 01 | Enameled copper wire diameter | 17 | µm |
| 02 | Number of turns | 30 | turns |
| 03 | Coil shape | Rectangular | ------ |
| 04 | Coil length | 1 | cm |
| 05 | Coil width | 1 | mm |
| 06 | Coil resistance after mounting on the sensor housing. | 83 | Ω |

A cantilever beam shape was cut out from a sheet of cellulose acetate (Overhead transparency Projector sheet: OHP sheet). The specifications of the cantilever are given in table-2. The selection of the material for the cantilever was based on its availability, low cost and ease of processing. The one end mounted cantilever arrangement was selected as it can bend under the action of impacting rain droplets and wind. The momentum transfer between the fluids and the cantilever should be highly efficient and at the same time the transferred energy from one form to another should be measureable, hence the particular material and the dimension was selected for the cantilever.

**Table-02:** Technical specifications of the cantilever

| Sr. no. | Description | Value | Unit |
|---|---|---|---|
| 01 | Cellulose acetate (Overhead transparency projector sheet) | ------ | ------ |
| 02 | Length 1 | 1.6 | cm |
| 03 | Length 2 | 1.2 | cm |
| 04 | Width | 1 | mm |
| 05 | Thickness | 100 | µm |

Figure 1(b) and (c) shows the image of the actual sensor and the cantilever-coil holder fabricated in laboratory for the testing purposes. Figure 1(b) shows the assembly of the coil which is suspended 2mm away from the surface of a high power rare earth magnet. Figure 1(c) shows the size of the coil compared with that of the insulin needle which is 313µm in diameter and having length of 1.6cm. The output of the coil was directly collected through the wires connected across the phoenix connector. Agilent multimeter 34401a was used to acquire all the voltages from the coil as it was exposed to droplet impact and wind flows respectively.

**Droplet impact testing setup of the sensor**
The sensor module was kept under the droplet dispensing tower by manually aligning them together. Droplets were dispensed from various heights and from different nozzles on to the sensor. The testing was carried out inside the laboratory. The dispenser had three different easy to change droplet dispensing nozzles. Table-3 provides the details about the three nozzles.

**Table-03:** Details of the nozzle types and their technical specifications

| Sr.no. | Nozzle type | Diameter of dispensed water droplet. | Unit |
|---|---|---|---|
| 01 | Insulin syringe needle | 1.9 | mm |
| 02 | 20ml syringe needle | 2.8 | mm |
| 03 | 200µl Pipette tip was cut at its dispensing end. (ID = 0.9mm and OD = 1.7mm) | 3.7 | mm |

**Turbulent wind speed sensor testing setup**
The sensor module was used without making any changes in its materials or the geometry for the sensing of wind turbulence. The sensor was directly exposed to a table fan driven wind flow which is in nature non laminar. The wind flow was directed to the sensor such that the flow was perpendicular to the page containing the figure 1(b). The velocity of the wind was simultaneously measured by a standardized thermoanemometer from TESTO, model no. 405i which was kept in the proximity of the cantilever-coil sensor.

**Natural Frequency of the cantilever-coil arrangement**
The cantilever with the coil was used with two different cantilever lengths. The natural frequencies of the two lengths were measured by using Analog Discovery circuit kit through MATLAB coding. The cantilever was imparted with forced oscillations through manual stimulus. As the free oscillations were induced in the cantilever-coil system, the induced voltage in the coil was logged into Analog discovery through Matlab programming. The same programme was able to provide FFT analysis of the acquired signal. Table-04 provides the details of the two natural frequencies for the two lengths of the cantilever.

**Table-04:** Physical length of the cantilever and the corresponding natural frequencies.

| Sr.no. | Cantilever length (cm) | Natural frequency (Hz) |
|---|---|---|
| 01 | 1.6 | 28.93 |
| 02 | 1.2 | 41.50 |

**Results and Discussions**

**Droplet impact testing of the sensor:** The sensor module was exposed to droplet impact and wind flow independently. Initially the sensor was used to study the response of impacting water droplets. Figure 2(a) and 2(b) shows the sensor response towards the impacting droplets of three diameters falling from various heights for two lengths of the cantilever.

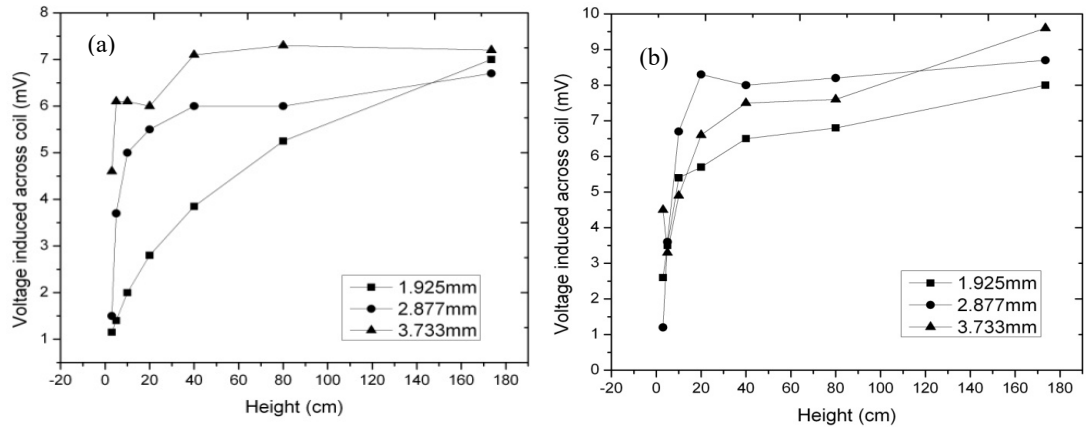

**Figure 2:** Induced voltage in coil for three different droplets falling from consecutively increasing heights. (a) Cantilever length 1.6cm (b) Cantilever length 1.2cm.

For a droplet falling from a height 'h' will have a maximum kinetic energy at the time of impact equal to product 'mgh' (mass of the droplet, acceleration due to gravity and height) if it is travelling at the terminal velocity. But in the laboratory testing facility the maximum available height was 173.5cm. As per previous reports [7-8], for droplets in the diameter range of 3-6mm to attained terminal velocity they need to traverse atleast 12-17m of length. As such the droplets which were used in the current work were not falling at terminal velocities. This implies that the momentum transfer by the droplets to the sensor coil should provide induce AC voltage corresponding to its final velocity at the time of impact. As size and fall length increases so should be the voltage. But the observe fact differs in this because of the natural frequency of the cantilever. The graph in the figure 2(a) corresponds to the cantilever with length 1.6cm and having natural frequency of 28.93Hz. That is roughly in 34 millisecond the cantilever executes its one full oscillation. This time period is still insufficient. This is because all of the droplets kinetic energy is imparted to the cantilever in less than that time period. As such the cantilever is incapable of reaching its full oscillation length in the time instant of the impact before the energy is dissipated in other forms. Therefore droplets falling from 60cm and above, even if they have greater momentum to transfer to the cantilever-coil system at the time of impact, it won't happen with 100% efficiency. This is evident from the fact that when the length of the cantilever was 1.2cm, the magnitude of the induce voltage for all the droplets increase significantly. This is because the natural frequency for this length is 41.50Hz which has time period equal to 24 milliseconds. This time period value is significantly less than the previous value but not small enough to achieve 100% efficiency. Hence, the cantilever with length of 1.2cm was able to provide more voltage as rate of change of flux through the coil was greater than it was when the length was 1.6cm respectively. The sensor can be further tuned for higher natural frequencies just by reducing its length. This way true momentum transfer of the droplets and subsequently their size can be obtained accurately. Another discrepancy that we can observe in the graphs of figure 2 is for a given size of the droplet, the magnitudes of the voltages are same for different heights. Referring to figure 1(a), there are two regions on the cantilever surface marked by number 1 and 2. If the droplet impacts on the region 1 we will get maximum amplitude for the cantilever oscillation. Likewise if the impact happens anywhere in between regions 1 and 2, the cantilever will oscillate with lower amplitude. It is very critical that the testing area must be free of any type of air flows. While the testing of this sensor was conducted, droplets tend to land on the cantilever at different positions as their dispensing height is increased. This is because some uncontrolled small disturbance in the air surrounding the experimental setup was flowing. This causes a big enough deflection in the final impacting point of the droplet on the cantilever. For the sensor to work in the field reliably, only the cantilever tip surface can be exposed to the falling rain droplets so that exact amplitude can be attained by the cantilever.

**Turbulent wind speed sensor testing:** The sensor module was used without any alterations in its materials or the geometry for the detection of the wind turbulence.

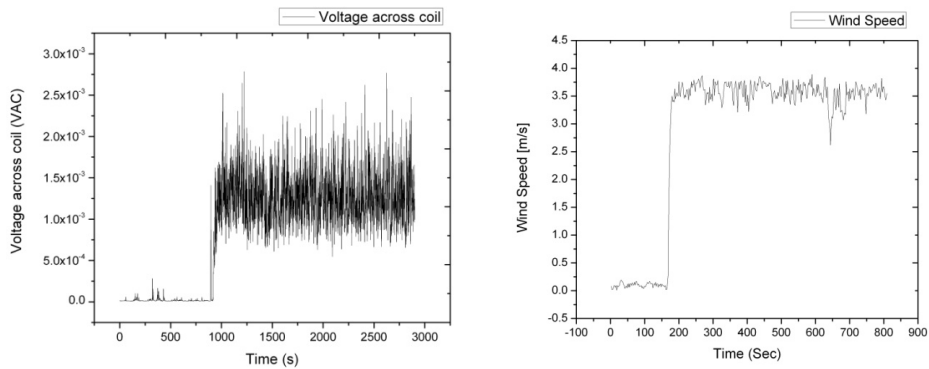

Figure 3: Sensors were exposed to wind coming from a table fan kept at a distance of 30cm. (a) Voltage induce in the coil for the cantilever length of 1.6cm. (b) Testo thermoanemometer wind velocity data.

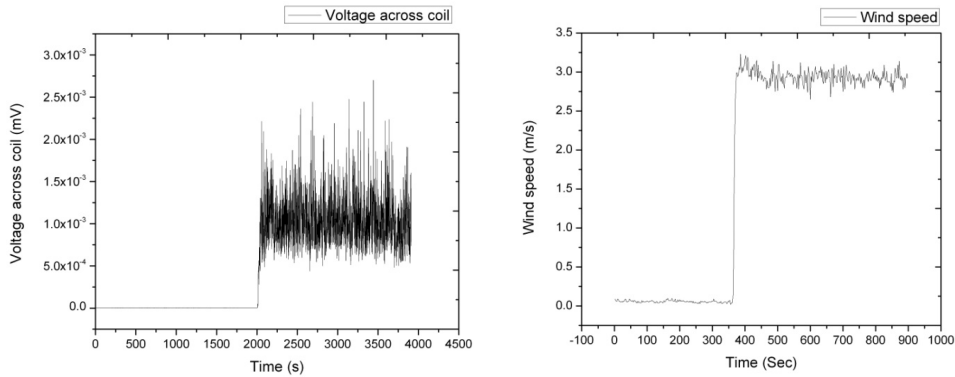

Figure 4: Sensors were exposed to wind coming from a table fan kept at a distance of 30cm. (a) Voltage induce in the coil for the cantilever length of 1.2cm. (b) Testo thermoanemometer wind velocity data.

It can be easily seen that from the figure 3 and 4, the cantilever-coil based sensor for both the oscillating lengths, it shows exactly similar trend in the induce voltage as compare to the data generated by the testo thermoanemometer sensor. Greater the turbulent intensity in a flow greater will be the voltage magnitude at that instant. This is what we can observe in the figure 4(a). The large spikes in the voltage values indicates the instances when the turbulent intensity in the wind flow was highest near the sensor. The flat line in the starting instants shows almost zero voltage which indicates that there is no wind flow as the fan was switched off. The coil produces voltage when it experiences change in flux which can happen only if the wind flow is unsteady (turbulent) which makes the cantilever oscillate. Based on just the oscillating length of the cantilever, the sensor can be tuned to numerous values of the natural oscillating frequecies which in turn can provide substantially accurate data of any turbulent wind flow velocity when the sensor is exposed to it.
By implementing proper housing to the sensor module, a good disdrometer or a standalone turbulence sensor can be obtained from the same single fabricated sensor.

**Conclusion**
A hybrid sensor based on principles of electromagnetic induction and response of cantilever to impulses was design and fabricated. The sensor was used to detect two different physical parameters without making any changes in its materials used or in its geometry. The dirt cheap cost of the fabrication of a single sensor module makes it most useful in sensing rainfall droplet size distribution by deploying numerous such sensors/sensor arrays over a very large area. The impulsive nature of the droplet impact creates substantially bigger magnitude of induce voltage as compare to the induce voltage by the turbulent wind flow. This makes the sensor useful in real time application to carry out measurement of both the physical quantities simultaneously with the supplementary use of simple electronics instrumentation. Practically most of the problems were solved within the scope of the work. Thus the feasibility study of the developed sensor shows promising results for implementation of this sensor for its respective use in the field after its calibration.

**Acknowledgement**
The authors wish to thank Prof. S. AnanthaKrishnan, Adjunct Professor and Raja Ramanna Fellow, Department of Electronics Science, Savitribai Phule Pune University (SPPU) for providing his laboratory and the resources for this work. Authors are thankful towards Mr. Mukesh Khanore, Physics department, (SPPU) for his technical and motivational discussions for this work.